\documentclass[journal=jacsat,manuscript=article]{achemso}

\newcommand{\figurescale}{1}
\usepackage{verbatim}
\usepackage{amsmath}
\usepackage[utf8]{inputenc}
\usepackage[separate-uncertainty=true]{siunitx}
\DeclareSIUnit{\rpm}{rpm}
\usepackage[colorlinks=true,urlcolor=blue,linkcolor=blue,citecolor=blue]{hyperref}
\usepackage{xcolor}
\usepackage[T1]{fontenc}
\usepackage{placeins}
\usepackage{nicefrac}
\usepackage{braket}
\usepackage{pdfcomment}
\usepackage{xcolor}
\usepackage{braket}
\usepackage{lineno}
\usepackage{atbegshi}

\AtBeginDocument{\AtBeginShipoutNext{\AtBeginShipoutDiscard}}
\author{J.~Klein}\email{julian.klein@wsi.tum.de}
\affiliation{Walter Schottky Institut and Physik Department, Technische Universit\"at M\"unchen, 85748 Garching, Germany}
\alsoaffiliation{Nanosystems Initiative Munich (NIM), 80799 Munich, Germany}
\author{A.~Kerelsky}
\affiliation{Department of Physics, Columbia University, New York, New York 10027, United States}
\author{M.~Lorke}
\affiliation{Institut für Theoretische Physik, Universität Bremen, 28334 Bremen, Germany}
\alsoaffiliation{Bremen Center for Computational Materials Science, University of Bremen, 28359 Bremen, Germany}
\author{M.~Florian}
\affiliation{Institut für Theoretische Physik, Universität Bremen, 28334 Bremen, Germany}
\author{F.~Sigger}
\affiliation{Walter Schottky Institut and Physik Department, Technische Universit\"at M\"unchen, 85748 Garching, Germany}
\author{J.~Kiemle}
\affiliation{Walter Schottky Institut and Physik Department, Technische Universit\"at M\"unchen, 85748 Garching, Germany}
\author{M.~C.~Reuter}
\affiliation{IBM T. J. Watson Research Center, Yorktown Heights, NY 10598, United States}
\author{T.~Taniguchi}
\affiliation{National Institute for Materials Science, Tsukuba, Ibaraki 305-0044, Japan}
\author{K.~Watanabe}
\affiliation{National Institute for Materials Science, Tsukuba, Ibaraki 305-0044, Japan}
\author{J.~J.~Finley}
\affiliation{Walter Schottky Institut and Physik Department, Technische Universit\"at M\"unchen, 85748 Garching, Germany}
\alsoaffiliation{Nanosystems Initiative Munich (NIM), 80799 Munich, Germany}
\alsoaffiliation{Munich Center for Quantum Science and Technology (MCQST), 80799 Munich, Germany}
\author{A.~Pasupathy}
\affiliation{Department of Physics, Columbia University, New York, New York 10027, United States}
\author{A.~W.~Holleitner}\email{holleitner@wsi.tum.de}
\affiliation{Walter Schottky Institut and Physik Department, Technische Universit\"at M\"unchen, 85748 Garching, Germany}
\alsoaffiliation{Nanosystems Initiative Munich (NIM), 80799 Munich, Germany}
\alsoaffiliation{Munich Center for Quantum Science and Technology (MCQST), 80799 Munich, Germany}
\author{F.~M.~Ross}\email{fmross@mit.edu}
\affiliation{IBM T. J. Watson Research Center, Yorktown Height, NY 10598, United States}
\alsoaffiliation{Department of Materials Science and Engineering, Massachusetts Institute of Technology, Cambridge, MA 02139, United States}
\author{U.~Wurstbauer}\email{wurstbauer@uni-muenster.de}
\affiliation{Walter Schottky Institut and Physik Department, Technische Universit\"at M\"unchen, 85748 Garching, Germany}
\alsoaffiliation{Nanosystems Initiative Munich (NIM), 80799 Munich, Germany}
\alsoaffiliation{Institute of Physics, University of M\"unster, 48149 M\"unster, Germany}
\title{Impact of intrinsic and extrinsic imperfections on the electronic and optical properties of MoS$_2$}

\begin{document}

%

\begin{abstract}
\textbf{Intrinsic and extrinsic disorder from lattice imperfections, substrate and environment has a strong effect on the local electronic structure and hence the optical properties of atomically thin transition metal dichalcogenides that are determined by strong Coulomb interaction. Here, we examine the role of the substrate material and intrinsic defects in monolayer MoS$_{2}$ crystals on SiO$_{2}$ and hBN substrates using a combination of scanning tunneling spectroscopy, scanning tunneling microscopy, optical absorbance, and low-temperature photoluminescence measurements.  We find that the different substrates significantly impact the optical properties and the local density of states near the conduction band edge observed in tunneling spectra. While the SiO$_{2}$ substrates induce a large background doping with electrons and a substantial amount of band tail states near the conduction band edge of MoS$_{2}$, such states as well as the high doping density are absent using high quality hBN substrates. By accounting for the substrate effects we obtain a quasiparticle gap that is in excellent agreement with optical absorbance spectra and we deduce an exciton binding energy of about 480~meV. We identify several intrinsic lattice defects that are ubiquitious in MoS$_{2}$, but we find that on hBN substrates the impact of these defects appears to be passivated. We conclude that the choice of substrate controls both the effects of intrinsic defects and extrinsic disorder, and thus the electronic and optical properties of MoS$_{2}$. The correlation of substrate induced disorder and defects on the electronic and optical properties of MoS$_{2}$ contributes to an in-depth understanding of the role of the substrates on the performance of 2D materials and will help to further improve the properties of 2D materials based quantum nanosystems.} 
\end{abstract}

%
\maketitle
%
%
\section{Introduction}

The electronic structure of atomically thin transition metal dichalcogenides (TMDCs) is unique due to the weak non-local dielectric screening of the Coulomb interaction.~\cite{Chernikov.2014} The large surface-to-volume ratio makes these materials highly sensitive to changes in the dielectric environment. This has been exploited as a tuning knob to control excitonic properties and quasiparticle gaps in these materials.~\cite{Raja.2017,Florian.2018} Taking full advantage of TMDCs requires an understanding of the relationship between the electronic properties of the materials and their environment. This involves measuring key electronic and optical properties in the presence of both extrinsic effects from the surroundings and intrinsic, atomic scale defects. The electronic properties of TMDCs can be probed by various experimental methods. Angle-resolved photo emission spectroscopy (ARPES), scanning tunneling spectroscopy (STS), photoluminescence excitation (PLE) spectroscopy and optically probing Rydberg states allow the determination of the quasiparticle band gap E$^{qp}_ {gap}$ energy.~\cite{Zhang.2014,Klots.2014,Chiu.2015,Hill.2015,Huang.2015,Rigosi.2016,Zhou.2016,Hill.2016,Yao.2017,Kerelsky.2017,Robert.2018} The values are typically compared to the onset energy E$_{opt}$ in optical absorbance or photoluminescence spectra to obtain the excitonic binding energy E$_{bin}$ = E$^{qp}_ {gap}$ - E$_{opt}$. Experimental~\cite{Zhang.2014,Klots.2014,Chiu.2015,Hill.2015,Rigosi.2016,Zhou.2016,Hill.2016,Yao.2017,Kerelsky.2017,Robert.2018} and theoretical~\cite{Ramasubramaniam.2012,Qiu.2013,Cheiwchanchamnangij.2012,Shi.2013,Steinhoff.2014,Liang.2015,Florian.2018} values for the binding energy vary widely in the literature, even for similar device geometries. In general, scanning tunneling spectroscopy (STS) is the method of choice for obtaining the quasiparticle gap from tunneling spectra since it is a direct probe of the local density of states (LDOS). Scanning tunneling microscopy (STM) is a favourable tool for visualizing defects with sub-$\SI{}{\nano\meter}$ spatial resolution and determining their densities in mono- and few-layer TMDCs and other materials.~\cite{McDonnell.2014,Addou.2015,Addou.2015b,Liu.2016,Vancso.2016,Park.2017,Coelho.2018,DelaMarion.2018} 

Transition metal dichalcogenide (TMDC) crystals are often prepared by mechanical cleavage and then transferred onto SiO$_{2}$ substrates that allow optical identification. However, SiO$_{2}$ as a substrate is well-known to exhibit dangling bonds and near-surface charge traps and is therefore not the substrate of choice for applications that require good sample quality and homogeneity.~\cite{Dolui.2013,Zhu.2014} In contrast, the use of hBN as a substrate in van der Waals heterostructures has been widely established since it significantly enhances the electronic and photo-physical properties of atomically thin TMDCs, greatly reducing inhomogeneous linewidth broadening in photoluminescence ~\cite{Wierzbowski.2017,Ajayi.2017,Cadiz.2017,Florian.2018} and enabling large carrier mobilities due to strongly reduced scattering in field effect devices.~\cite{Cui.2015,Dean.2010}

Here, we combine STS, optical absorbance and photoluminescence spectroscopy to investigate the impact of these two different substrates on the electronic and optical properties in monolayer MoS$_{2}$. We complementary measure the structure and density of intrinsic atomic scale defects. We quantitatively compare STS spectra recorded on exfoliated MoS$_{2}$ monolayers placed on SiO$_{2}$ and hBN. Using SiO$_{2}$ as substrate, we find significant contribution of band tail states (BTS) below the conduction band (CB) that obscure the determination of the quasiparticle gap E$^{qp}_ {gap}$ and in turn the extracted exciton binding energy E$_{bin}$. The BTS are absent when using hBN as a substrate. Moreover, the use of hBN reduces the electron density shifting the Fermi level further to the center of the band gap, making the TMDC crystal more intrinsic. We find that the n-type doping is, due to both the SiO$_{2}$ substrate and intrinsic defects. The later most likely due to a large density of native sulfur vacancies that we determine from STM topography measurements and in agreement with previous reports.~\cite{Vancso.2016,Qiu.2013b,Hong.2015}  Moreover, low-temperature photoluminescence spectroscopy on nominally identical samples and on fully hBN encapuslated MoS$_{2}$ monolayers is performed to cross-correlate the influence of the substrate on the electronic and optical properties. Taking into accound the various spectroscopy methods applied to the different sample geometries we discuss that the impact of the  environment and SiO$_{2}$ substrates on the intrinsic lattice defects are passivated if an hBN substrate or encapsulation with hBN is used. 

\section{Results and discussion}

%
\begin{figure}[!ht]
\scalebox{\figurescale}{\includegraphics[width=0.7\linewidth]{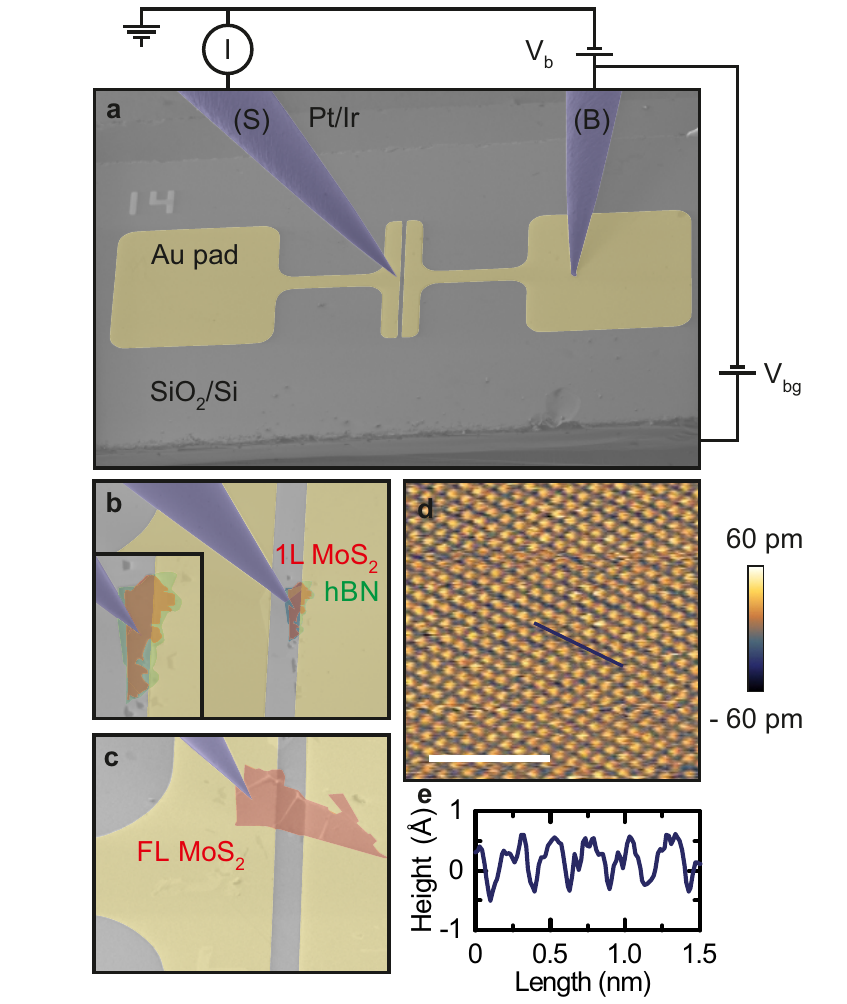}}
\renewcommand{\figurename}{Fig.}
\caption{\label{fig1}
\textbf{Room temperature STM of MoS$_{2}$ with atomic resolution.} 
\textbf{a}, SEM image of a gate-tunable sample in the STM chamber. The tunneling current is measured between scanning (S) and biasing (B) tips for a chosen bias voltage V$_{b}$. A backgate voltage V$_{bg}$ between biasing tip and the degenerately doped Si can be applied for tuning charge carrier densities in the sample.
\textbf{b}, Magnified SEM image of a van der Waals heterostructure consisting of atomically thin MoS$_{2}$ on hBN.
\textbf{c}, Magnified SEM image of few-layer MoS$_{2}$.
\textbf{d}, Typical constant current STM topography image of a defect-free region from the sample shown in \textbf{c} revealing atomic resolution.
\textbf{e}, Corresponding line cut in \textbf{d} showing interatomic S-S distance. 
}
\end{figure}

The samples investigated throughout this manuscript are fabricated by mechanical exfoliation and deterministic transfer onto thermally grown SiO$_{2}$/Si substrates or on few layer hBN flakes that are also supported by SiO$_{2}$/Si. The MoS$_{2}$ flakes are either contacted by direct transfer onto a thin $\SI{30}{\nano\meter}$ Au/$\SI{5}{\nano\meter}$ Ti pad or by evaporation of Au/Ti with similar thicknesses through a shadow mask to exclude any surface contamination from residues of photo-resist. STM/STS measurements are taken at room temperature. The STM is equipped with multiple tips and a scanning electron microscope (SEM) that allows mono- and few-layer MoS$_{2}$ samples and van der Waals heterostructures to be located and probed.~\cite{Ji.2011}

Figure~\ref{fig1}a shows a false color SEM image of a typical sample. A bias is applied between the biasing (B) and the scanning (S) tip which are both made from PtIr and a bias between the bias tip and the degenerately doped Si substrate can be applied to gate the device. A close-up SEM image of a monolayer MoS$_{2}$ on hBN van der Waals heterostack and a few-layer MoS$_{2}$ sample is shown in Fig.~\ref{fig1}b and c. A typical constant current topography map of few-layer MoS$_{2}$ is shown in Fig.~\ref{fig1}d revealing atomic resolution and the hexagonal lattice symmetry from the topmost sulfur layer. A corresponding line cut is shown in Fig.~\ref{fig1}(e) yielding the expected S-S distance of $\SI{3.17}{\angstrom}$.~\cite{Mattheiss.1973}

%
\begin{figure*}[!ht]
\scalebox{\figurescale}{\includegraphics[width=1\linewidth]{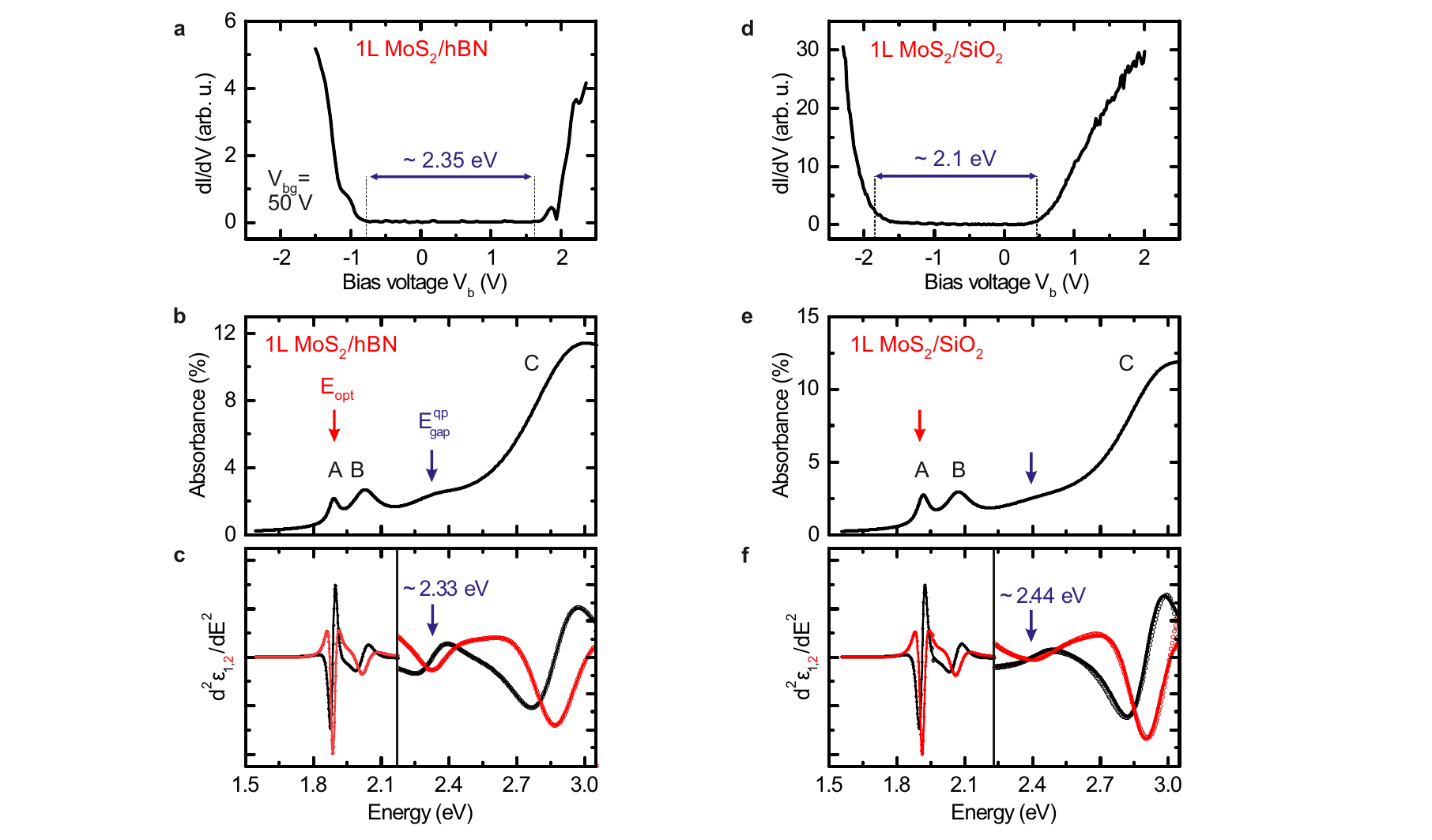}}
\renewcommand{\figurename}{Fig.}
\caption{\label{fig2}
\textbf{STS and optical absorbance of monolayer MoS$_{2}$ on hBN and SiO$_{2}$.} 
\textbf{a}, STS spectrum of monolayer MoS$_{2}$ on hBN taken at V$_{bg} = \SI{50}{\volt}$.
\textbf{b}, Optical absorbance of monolayer MoS$_{2}$ on hBN. The spectrum reveals the A and B exciton and the C resonance and the onset of the quasiparticle gap at $E^{qp}_{gap} = \SI{2.33}{\electronvolt}$.
\textbf{c} Critical point analysis of MoS$_{2}$ on hBN. The second derivative of the real (imaginary) part $\frac{d^{1(2)}\varepsilon_{2}}{dE^{2}}$ of the dielectric function are plotted as black (red) traces. Both reveal resonances at the A and B excitonic transitions as well as a weaker feature at $\SI{2.33}{\electronvolt}$ that is associated with E$_{gap}^{qp}$.
\textbf{d}, STS spectrum of monolayer MoS$_{2}$ on SiO$_{2}$ substrate.
\textbf{e}, Optical absorbance of monolayer MoS$_{2}$ on SiO$_{2}$. Besides the A and B exciton and the C resonance, the onset of the quasiparticle gap at E$_{gap}^{qp} = \SI{2.44}{\electronvolt}$ is also visible.
(f) Critical point analysis of MoS$_{2}$ on SiO$_{2}$. The second derivative of the real (imaginary) part $\frac{d^{1(2)}\varepsilon_{2}}{dE^{2}}$ of the dielectric function are plotted as black (red) traces. Both reveal resonances at the A and B excitonic transitions as well as weaker signature at $\SI{2.44}{\electronvolt}$ that is associated with E$_{gap}^{qp}$.
}
\end{figure*}

We now turn to the measurement of the local density of electronic states (LDOS) from STS and of the quasiparticle gap that is extracted from the LDOS. STS spectra $dI/dV$ that are proportional to the LDOS are shown in Fig.~\ref{fig2}a and d for monolayer MoS$_{2}$ on hBN and SiO$_{2}$ substrates, respectively. Vanishing values in the $dI/dV$ spectrum correspond to the absence of electronic states and allows the determination of an apparent gap in the LDOS. For ideal semiconducting materials, the onset of the $dI/dV$ for negative bias voltages can be assigned to the valence band edge (VB) and the onset for positive bias voltages the conduction band edge (CB). The position of the Fermi energy is at zero bias voltage. 

We begin with discussing the STS of monolayer MoS$_{2}$ stacked onto high quality hBN which is an atomically flat and charge neutral insulating 2D material.~\cite{Dean.2010} The $dI/dV$ spectrum of monolayer MoS$_{2}$ on hBN is shown in Fig.~\ref{fig2}a. In order to record the $dI/dV$ curve a backgate voltage of $V_{bg} = \SI{50}{\volt}$ is used to reach sufficient conductivity. The STS measurement shows an apparent gap in the LDOS of $\sim\SI{2.35}{\electronvolt}$ with both the CB and the VB exhibiting a steep increase in density of states. We perform optical absorbance spectroscopy on a nominally identical sample for direct comparison with the tunneling spectra. The optical absorbance spectrum is shown in Fig.~\ref{fig2}b. The spectrum reveals absorption from the A and B exciton transition at $\SI{1.89}{\electronvolt}$ and $\SI{2.05}{\electronvolt}$. Moreover, by performing a critical point analysis by converting the absorbance spectrum to the real and imaginary part of the dielectric function utilizing Kramers-Kronig constraint analysis~\cite{Li.2014} and taking the second derivative~\cite{Lautenschlager.1987,Yao.2017} as shown in Fig.~\ref{fig2}c, we determine a quasiparticle gap of $E^{qp}_{gap, Abs} \sim \SI{2.33}{\electronvolt}$, which is in excellent agreement with the LDOS gap obtained from tunneling spectra. Therefore, we assign the LDOS gap from STS to the quasiparticle gap $E^{qp}_{gap, STS} \sim \SI{2.35}{\electronvolt}$. From the emission of the 1s exciton (A exciton), we can directly determine the exciton binding energy through $E_{Bind} = E^{qp}_{gap} - E_{A}$. Taking $E^{qp}_{gap}$ from tunneling (absorbance) spectra we obtain an exciton binding energy of $\SI{460}{\milli\electronvolt}~(\SI{440}{\milli\electronvolt})$ in good agreement with theory that predicts values of $E_{Bind} = 300-\SI{700}{\milli\electronvolt}$~\cite{Ramasubramaniam.2012,Qiu.2013,Cheiwchanchamnangij.2012,Shi.2013,Liang.2015,Florian.2018} and binding energies obtained from optical spectroscopy measurements reporting of $E_{Bind} > \SI{600}{\milli\electronvolt}$.~\cite{Hill.2015,Yao.2017} 

Similar to the measurement on the hBN substrate, we perform tunneling and optical spectroscopy on monolayer MoS$_{2}$ on SiO$_{2}$ substrate as shown in Fig.~\ref{fig2}d and e, respectively. Here, the monolayer MoS$_{2}$ on the normally used SiO$_{2}$ substrate reveals an apparent gap in the LDOS of $\sim \SI{2.1}{\electronvolt}$, a value that is in good agreement with values from STS measurements on monolayer MoS$_{2}$ on SiO$_{2}$ in literature.~\cite{Zhang.2014,Chiu.2015,Rigosi.2016,Zhou.2016,Hill.2016,Kerelsky.2017} Intriguingly, we observe the onset of the quasiparticle gap from absorbance measurements in Fig.~\ref{fig2}e and the related critical point analysis in Fig.~\ref{fig2}f at an energy of $E^{qp}_{gap, Abs} \sim \SI{2.44}{\electronvolt}$, much higher than the value from the STS data. Determination of the binding energy from pure optical measurements shown in Fig.~\ref{fig2}e reveals $E_{Bind} = E^{qp}_{gap, Abs} - E_{A} = \SI{530}{\milli\electronvolt}$, taking into account the optical gap $E_{A} = \SI{1.91}{\electronvolt}$, in good agreement with theory and other optical measurements, but more than twice the binding energy using the apparent gap from STS measurements shown in Fig.~\ref{fig2}d.

In TMDCs, the exciton binding energy directly depends on the surrounding dielectric environment. For the static case and in the limit of long wavelengths ($q \rightarrow 0$), the dielectric constant of the environment is approximated by an average dielectric constant $\varepsilon_{ave} = \frac{\varepsilon_{top}+\varepsilon_{bottom}}{2}$ of the surrounding dielectrics.~\cite{Keldysh.1979,Cudazzo.2011} For air/vacuum)/MoS$_{2}$/SiO$_{2}$, we obtain an average dielectric constant of $\varepsilon_{ave} = \frac{\varepsilon_{air} + \varepsilon_{SiO_{2}}}{2} = 2.45$ with $\varepsilon_{SiO_{2}} = 3.9$.~\cite{Philipp.1979} For air(vacuum)/MoS$_{2}$/hBN, we get $\varepsilon_{ave} = 3.45$ with $\varepsilon_{hBN} = \sqrt{\varepsilon_{\bot}\varepsilon_{\parallel}} = 5.89$.~\cite{Geick.1966} Considering these values, we expect an exciton binding energy that is slightly larger for air(vacuum)/MoS$_{2}$/SiO$_{2}$ compared to air(vacuum)/MoS$_{2}$/hBN due to a weaker dielectric screening. The $\varepsilon_{ave}$ for air(vacuum)/MoS$_{2}$/hBN is most likely to be overestimated due to an air gap between MoS$_{2}$ and hBN of $\sim \SI{5}{\angstrom}$~\cite{Rooney.2017} that lowers dielectric screening.~\cite{Florian.2018} This explains why $E_{bin}$ is very similar in both dielectric environments as observed in the all optical absorbance measurements, but a few tens of $\SI{}{\milli\electronvolt}$ larger for the MoS$_{2}$ monolayer on SiO$_{2}$ with $E_{bin}\sim \SI{440}{\milli\electronvolt}$ and  $\sim \SI{480}{\milli\electronvolt}$ for the 1s exciton in MoS$_{2}$ on hBN and SiO$_{2}$, respectively. From similar considerations regarding the dielectric environment, we expect to obtain also very similar quasiparticle gaps $E^{qp}_{gap}$ for MoS$_{2}$ on SiO$_{2}$ and on hBN in agreement with the optical measurements but in disagreement with the apparent gap in the LDOS from STS measurements.

%
\begin{figure}[!ht]
\scalebox{\figurescale}{\includegraphics[width=0.6\linewidth]{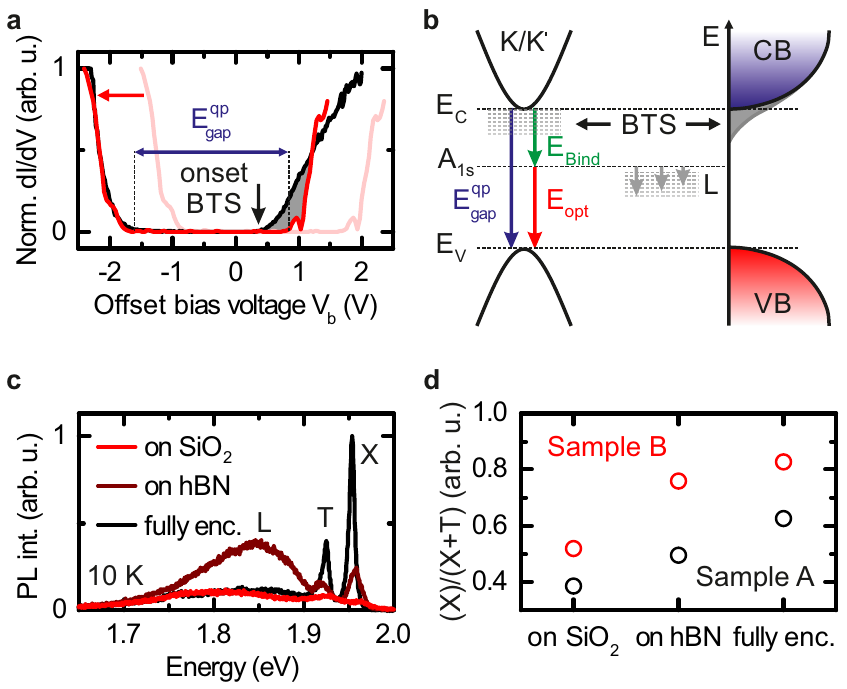}}
\renewcommand{\figurename}{Fig.}
\caption{\label{fig3}
\textbf{Emergence of band tail states and low-temperature ($\SI{10}{\kelvin}$) photoluminescence measurements of monolayer MoS$_{2}$ on hBN and SiO$_{2}$.} 
\textbf{a}, Normalized STS spectra of monolayer MoS$_{2}$ on SiO$_{2}$ and hBN for direct comparison. Contribution from band tail states (BTS) in the conduction band minimum (CBM) on SiO$_{2}$ are highlighted.
\textbf{b}, Schematic of the interplay of quasiparticle gap $E^{qp}_{gap}$, optical gap $E_{opt}$ and exciton binding energy $E_{bind}$. The presence of BTS gives rise to density of electronic states tailing off the conduction band edge. 
\textbf{c}, Low-temperature ($\SI{10}{\kelvin}$) $\mu$-PL of monolayer MoS$_{2}$ on SiO$_{2}$, on hBN and fully encapsulated between hBN. The spectrum shows emission from neutral and charged excitons as well as the defect related $L$-band.
\textbf{d}, Spectral weight of the neutral and charged exciton for the corresponding sample geometries in \textbf{c} obtained from two nominally identical samples.
}
\end{figure}

To further investigate the origin of this discrepancy, we directly compare the STS spectra for MoS$_{2}$ monolayers on SiO$_{2}$ and hBN. We make two main observations: First, it is evident that in addition to the larger STS gap, the Fermi level is shifted close to the middle of the apparent gap in the LDOS with hBN as substrate compared to SiO$_{2}$. For MoS$_{2}$ on SiO$_{2}$, the apparent gap is smaller and the Fermi level is close to the conduction band edge, showing the commonly observed n-type behavior of MoS$_{2}$. The origin of excess electrons is not yet fully understood, but has been tentatively attributed to defects in the crystal lattice~\cite{Noh.2014} and charge traps in the SiO$_{2}$ substrate.~\cite{Dolui.2013} Second, the rise of LDOS at the conduction band edge is much steeper for monolayers on hBN than on SiO$_{2}$ (see Fig.~\ref{fig2}a,d). The rise of the LDOS at the valence band edge is similar on both substrates. This is further visualized in Fig.~\ref{fig3}a by overlapping the onset for negative bias voltages in the $dI/dV$ traces assigned to the valence band maxima by offsetting the curves to match the valence bands for MoS$_{2}$ on both substrates. The LDOS clearly matches in the valence band tail but is strictly steeper at the onset, attributed to differences in the conduction band for MoS$_{2}$ on hBN than on MoS$_{2}$ on SiO$_{2}$. 

We attribute the additional and more slowly raising LDOS at the CB of MoS$_{2}$ on SiO$_{2}$ to the emergence of band tail states (BTS) due to the SiO$_{2}$ substrate (see Fig.~\ref{fig3}b).~\cite{Zhu.2014} The surface of SiO$_{2}$ is subject to dangling bonds and near-surface charge traps~\cite{Dolui.2013} that can severely impact the electronic properties of TMDCs as observable in tunneling spectroscopy by the introduction of interfacial states between the SiO$_{2}$ and the MoS$_{2}$. In particular, silanol groups (Si-O-H) are expected at the SiO$_{2}$ surface due to $O_{2}$ plasma treatment prior to the MoS$_{2}$ transfer.~\cite{Nagashio.2011} The silanol groups are negatively charged and are reported to exist in large densities of $5 \cdot 10^{14} \SI{}{\per\centi\meter\squared}$.~\cite{Nagashio.2011} Similarly, it has been reported for electronic devices prepared from CVD grown MoS$_{2}$ on SiO$_{2}$ that the electronic transport properties are severely limited by such band tail trapping states.~\cite{Zhou.2013} Earlier work also described the influence of surface charge traps on the transport properties of MoS$_{2}$.~\cite{Ayari.2007}. As a consequence of the existence of such substrate induced BTS, the onset of LDOS in STS measurements for positive bias voltages does not directly mark the conduction band minimum. The onset often referred to as the CB edge is obscured by the additional trapping states tailing off from the conduction band edge giving the impression of a lowered $E^{qp}_{gap}$ from STS measurements. Since the emission and absorption energy of the exciton stays unaffected, an underestimated exciton binding energy is derived. This is sketched in Fig.~\ref{fig3}b.

To further investigate the difference in the apparent gap in the LDOS on the SiO$_{2}$ substrate and the impact of the substrate induced imperfections on the optical properties, we perform low-temperature ($\SI{10}{\kelvin}$) photoluminescence measurements of monolayer MoS$_{2}$ on SiO$_{2}$, on hBN and for comparison with reports in literature also on fully encapsulated in hBN (see Fig.~\ref{fig3}c). The spectra exhibit emission from the neutral exciton $X$, the negatively charged trion $T$ and the $L$ peak that is attributed to adsorbates and intrinsic defects. Change from SiO$_{2}$ to hBN substrate already significantly narrows down the free exciton linewidths as recently demonstrated.~\cite{Wierzbowski.2017,Ajayi.2017,Cadiz.2017,Florian.2018} Moreover, the narrowing is accompanied by a significant reduction of the $L$ peak and a slight shift of spectral weight from trion to the neutral exciton indicating that the MoS$_{2}$ crystal is less electron doped and, therefore, more intrinsic when hBN replaces SiO$_{2}$. This is in good agreement with the existence of silanol groups at the SiO$_{2}$ that result in unintentional n-type doping.

The reduced electron doping for monolayers on hBN is at least partially a very likely explanation for the observation that in STS spectroscopy the Fermi level shifts by almost $\SI{1}{\electronvolt}$ closer to the valence band compared to monolayers on SiO$_{2}$ as is evident in Fig.~\ref{fig2}a,d and Fig.~\ref{fig3}a. We would like to note that an additional contribution to this large shift in the Fermi energy could be caused by a reduction of the work function of the MoS$_{2}$ on hBN, an effect that is widely observed for hBN/metal heterostructures.~\cite{Morscher.2006,Joshi.2012,Schulz.2014,Zhang.2016,Zhang.2018,Parzinger.2017,Diefenbach.2018}

We further quantify the interpretation that the doping is reduced for Mo2$_{2}$ on hBN by plotting the relative spectral weight of the neutral exciton $\frac{X}{X+T}$ in Fig.~\ref{fig3}d for two nominally identical samples. For both samples we observe that the crystal is more intrinsic when stacked or even fully sandwiched in hBN. Moreover, photoluminescence of the L peak is reduced for hBN as substrate and even further when fully encapsulated. The origin of L peak emission and intrinsic doping can be manifold but is likely due to a complex interplay of substrate, physi- and chemisorbed atoms or molecules and the presence of native defects. In particular, the native defects are likely to interact with the substrate and/or physi- and chemisorbed atoms and molecules at the exposed surface. Interestingly, for MoS$_{2}$ on SiO$_{2}$, the energy difference between the onset of the L peak and the neutral exciton $X$ in low-temperature photoluminescence is about $\SI{200} {\milli\electronvolt}$ (compare Fig.~\ref{fig3}c and schematic in Fig.~\ref{fig3}b). This difference is almost identical to the energy difference between the onset of the density of states in STS spectroscopy interpreted as BTS and the actual conduction band edge as determined from absorbance measurements as well as to the onset at positive bias voltages in the STS spectra taken on MoS$_{2}$ on hBN (see Fig.~\ref{fig2}a). The similar energy differences suggest that the BTS in STS and the L peak in photoluminescence have the same or very similar origin.

%
\begin{figure}[!ht]
\scalebox{\figurescale}{\includegraphics[width=0.6\linewidth]{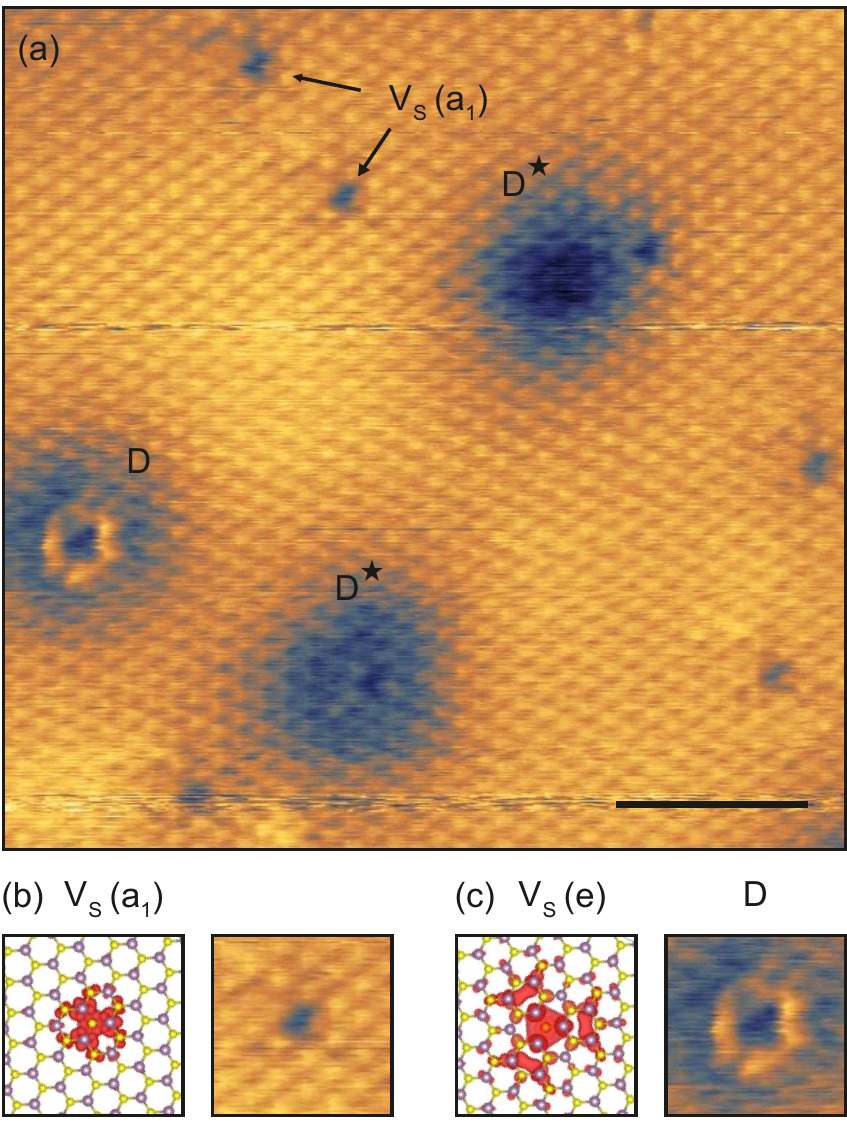}}
\renewcommand{\figurename}{Fig.}
\caption{\label{fig4}
\textbf{STM topography of native defects in few-layer MoS$_{2}$ at $V_{b} = \SI{1.2}{\volt}$ and $I = \SI{90}{\pico\ampere}$.}
\textbf{a}, Constant current STM topography of few-layer MoS$_{2}$ at room-temperature. The neutral mono-sulfur vacancy V$_{S}^{0}$ is ubiquitious. A second prominent defect $D$ with a larger wave function extent is also observed in addition to blue patches that most likely resemble defects $D^{\star}$ from the layer below. The scale bar is $\SI{2}{\nano\meter}$.
\textbf{b}, Wave function of the V$_{S}$ singlet state, $a_{1}$, from ab-initio calculations.
\textbf{c}, Ab-initio calculated wave function of the V$_{S}$ doublet state, $e$ compared to defect $D$ in experiment.
}
\end{figure}

In order to gain an in-depth understanding of the role of intrinsic imperfections on the optical and electronic properties, we record constant current STM topography maps on exfoliated MoS$_{2}$ to identify and quantify the type and density of intrinsic defects in our crystals. Here, we confine our investigations to few-layer MoS$_{2}$ allowing us to resolve defects in the top layer even at room temperature. A typical STM topography scan is presented in Fig.~\ref{fig4}a for positive bias. Our data reveals two prominent types of defects that are observed in all recorded STM maps on all investigated samples and areas. The most abundant defect is interpreted as single sulfur vacancy $V_{S}$ that resembles a missing sulfur atom on the top plane of the crystal. The spatial extent of the wave function is small ($<\SI{5}{\angstrom}$) coinciding well with the corresponding ab-initio calculated wave function of the singlet $a_{1}$ state of $V_{S}$ (see Fig.~\ref{fig4}b). This defect occurs with a large intrinsic defect density of $(4.6\pm 1.0)\cdot10^{12}\SI{}{\per\centi\meter\squared}$. This value is up to almost one order of magnitude lower than suggested by TEM studies on exfoliated MoS$_{2}$~\cite{Qiu.2013b,Hong.2015} but agrees with the lower bound given in another recent STM study.~\cite{Vancso.2016} The other defect that is consistently observed has a much larger wave function extent ($\sim\SI{15}{\angstrom}$) that is spread over a couple of interatomic S-S distances. In a previous STM study, this defect has been attributed to the doublet $e$ state of the mono-sulfur vacancy.~\cite{Vancso.2016} We are not entirely convinced by this attribution due to the very large defect size that does not coincide with ab-initio calculated wave functions (see Fig.~\ref{fig4}c). This defect shows up with a density of $(1.5\pm 0.5)\cdot10^{12}\SI{}{\per\centi\meter\squared}$. The blue patches in Fig.~\ref{fig4}a are very likely other defects ($D^{\star}$) that are either charged or in a layer below the surface.

The density of sulfur vacancies is in good agreement with the presence of negatively charged trions in photoluminescence by taking into account one excess electron per vacancy resulting in the intrinsic electron doping of $>10^{12}\SI{}{\per\centi\meter\squared}$ in good agreement with values found in the literature for MoS$_{2}$.~\cite{Miller.2015} Comparing the intrinsic defect density in light of STS measurements on both substrates, we notice that the hBN apparently passivates the impact of such defects resulting in a reduced electron doping visible by a Fermi energy that is shifted more closely to the center of the band gap in STS, and by a reduced trion and L peak emission in photoluminescence measurements. The overall effect is very similar to sulfur vacancy passivation through super-acid treatment.~\cite{Amani.2015,Cadiz.2016} We find it significant that the impact of defects appears to be passivated by the hBN substrate, since the STM imaging shows exfoliated MoS$_{2}$ is likely to have a high intrinsic defect density. When MoS$_{2}$ is interfaced to unpoassivated surfaces like SiO$_{2}$ or air, atoms and molecules are likely to be physi- and chemisorbed to native defects acting e.g. as molecular gates \cite{Miller.2015} changing the charge carrier density and the local electronic and optical properties. In our interpretation, intrinsic and extrinsic imperfections cause the experimentally observed BTS, a larger inhomogeneous broadening of excitonic interband transitions in photoluminescence, the L peak emission and an intrinsic n-type doping. In contrast, using hBN substrates allows the unambiguous identification of the quasiparticle gap unobscured by any BTS and in addition to narrow exciton emission with reduced intrinsic doping and suppressed L peak emission when fully encapsulated.

\section{Conclusion}

We have investigated the electronic properties of mono- and few-layer MoS$_{2}$ on different substrates and the influence of extrinsic and intrinsic imperfections. SiO$_{2}$ induces BTS below the conduction band minimum, which lowers the LDOS gap measured by STS for monolayer MoS$_{2}$. In contrast, hBN substrates are superior in quality, resulting in very clean tunneling spectra and high quality absorbance and photoluminescence spectra indicating the absence of defect and trapping states tailing off the conduction band edge. We find a high intrinsic density of sulfur atoms vacancies that is likely to impact the electronic properties by interaction with substrate and physi- and chemisorbed atoms and molecules. The deteriorating influence of extrinsic and intrinsic imperfections can be mitigated by proper choice of the substrate or by fully encapsulating the MoS$_{2}$, e.g. in hBN. These results help to understand and control the influence of substrates and defects on the electronic structure of TMDCs.

\section{Methods}
\subsection{Sample structure}
We employed the viscoelastic transfer method~\cite{CastellanosGomez.2014} to transfer MoS$_2$ monolayer crystals cleaved from bulk MoS$_{2}$ (SPI Supplies) and hBN multilayers onto $\SI{290}{\nano\meter}$ SiO$_2$ substrates. The hBN bulk crystals are provided by Takashi Taniguchi and Kenji Watanabe from NIMS, Japan.

\subsection{Scanning tunneling microscopy and spectroscopy}
To obtain clean surfaces for STM, samples are vacuum annealed at about $\SI{400}{\kelvin}$ for $\SI{600}{\second}$ under a vacuum of $10^{-6}\SI{}{}$ torr in the loadlock of the microscope. The samples are then transferred into an ultra-high vacuum chamber at $2\cdot 10^{-10}\SI{}{}$ torr that contains a four-probe scanning tunneling microscope.~\cite{Ji.2011} The microscope is operated either for recording constant current topographies or constant height transfer characteristics. The latter are post-processed by differentiation in order to obtain $dI/dV$ spectra presented in the manuscript. All data are acquired with a PtIr tip that is prepared \textit{in situ} in the microscope by applying a high electric field while approaching a Au target. The STM is equipped with a scanning electron microscope that allows us to conveniently locate and probe mono- and few-layer van der Waals materials and their heterostructures.

\subsection{Optical spectroscopy}
The optical absorbance is determined by reflectance measurements of the samples at room temperature using broadband emission from a Xenon arc discharge lamp $\SI{150}{\watt}$ (LOT). The reflected light is collected with a 50x objective ($\text{NA}=0.5$, N Plan, infinity corrected) and spatially filtered by a $\SI{50}{\micro\meter}$-core multimode fiber. A spot size of $\SI{6}{\micro\meter}$ is obtained on the sample. The collected light is guided to a VIS-wavelength range spectrometer (Horiba VS140). We obtain the absorbance from the reflectance data using a Kramers-Kronig constrained variational analysis. The values for the quasi-particle band gap are extracted by applying a critical point analysis to the absorbance data.

For low-temperature photoluminescence spectroscopy the sample is kept in a He-flow cryostat at a lattice temperature of $\SI{10}{\kelvin}$. A CW laser with an excitation energy of $\SI{2.33}{\electronvolt}$ and an excitation power of $\sim\SI{10}{\micro\watt}$ is focused through a 100x microscope objective ($\text{NA}=0.55$) onto the sample with a spot size of $\sim\SI{1}{\micro\meter}$. The emitted light is collected with the same microscope objective and guided to a spectrometer. The light is dispersed on a grating and detected with a liquid nitrogen cooled charge-coupled device (CCD).\\

\subsection{Density functional theory}
The calculations were performed using density functional theory (DFT), with the projected augmented wave method, as implemented in VASP~\cite{VASP:3,VASP:4,Kresse:99,Bloechl}. The electronic structures were determined using the PBE exchange correlation functional. A plane wave basis with an energy cutoff of ${\rm E_{cut}=500\,eV}$ and a $(6\times6\times1)$ $\Gamma$-centered Monkhorst-Pack {\bf k}-point sampling was used. The TMD layer has been modeled using a $(9\times9)$ supercell containing 242 atoms with periodic boundary conditions.\\

%
%
\section{Acknowledgements}
Supported by Deutsche Forschungsgemeinschaft (DFG) through the TUM International Graduate School of Science and Engineering (IGSSE). We gratefully acknowledge financial support of the Deutsche Forschungsgemeinschaft (DFG, German Research Foundation) under Germany's Excellence Strategy – EXC 2089/1 – 390776260, EXC-2111 – 390814868, and the Nanosystems Initiative Munich as well as the PhD program ExQM of the Elite Network of Bavaria. M.L. was supported by the Deutsche Forschungsgemeinschaft (DFG) within RTG 2247 and through a grant for CPU time at the HLRN (Hannover/Berlin).\\

\section{Author contributions}
J.K., J.J.F., A.P., A.W.H., F.M.R., U.W. conceived and designed the experiments, J.K., F.S. and J.Ki. prepared the samples, K.W. and T.T.provided high-quality hBN bulk crystals, J.K., A.K. and M.C.R. performed scanning tunneling spectroscopy measurements, J.K. and J.Ki. performed the optical measurements, M.L. and M.F. performed DFT calculations, J.K. analyzed the data, J.K. wrote the manuscript with input from all coauthors. \\


%
\section{Additional information}

\subsection{Competing financial interests} The authors declare no competing financial interests.

%
%
\bibliographystyle{achemso}
\providecommand{\latin}[1]{#1}
\makeatletter
\providecommand{\doi}
  {\begingroup\let\do\@makeother\dospecials
  \catcode`\{=1 \catcode`\}=2 \doi@aux}
\providecommand{\doi@aux}[1]{\endgroup\texttt{#1}}
\makeatother
\providecommand*\mcitethebibliography{\thebibliography}
\csname @ifundefined\endcsname{endmcitethebibliography}
  {\let\endmcitethebibliography\endthebibliography}{}

\end{document}